\documentclass[prl,showpacs,twocolumn]{revtex4}
\usepackage{bm}
\usepackage{amsmath}
\usepackage{amssymb}
\usepackage{graphicx}
\usepackage{slashed}
\usepackage{accents}
\usepackage{mathrsfs}

\begin{document}
\title{Quantum electron self-interaction in a strong laser field}
\author{S. \surname{Meuren}}
\author{A. \surname{Di Piazza}}
\email{dipiazza@mpi-hd.mpg.de}
\affiliation{Max-Planck-Institut f\"ur Kernphysik, Saupfercheckweg 1, D-69117 Heidelberg, Germany}
\date{\today}

\begin{abstract}
The quantum state of an electron in a strong laser field is altered if the interaction of the electron with its own electromagnetic field is taken into account. Starting from the Schwinger-Dirac equation, we determine the states of an electron in a plane-wave field with inclusion, at leading order, of its electromagnetic self-interaction. On the one hand, the electron states show a pure ``quantum'' contribution to the electron quasi-momentum, conceptually different from the conventional ``classical'' one arising from the quiver motion of the electron. On the other hand, the electron self-interaction induces a distinct dynamics of the electron spin, whose effects are shown to be measurable in principle with available technology.

 
\end{abstract}
 
\maketitle

The predictions of Quantum electrodynamics (QED) have been demonstrated experimentally with extremely high accuracy. On the other hand, apart from numerous successful investigations on QED effects in highly-charged ions \cite{Stoehlker_2008}, relatively little is known experimentally about QED in the presence of strong background electromagnetic fields. Here ``strong'' means of the order of the so-called ``critical'' electric (magnetic) field of QED: $E_{cr}=m^2c^3/\hbar|e|=1.3\times 10^{16}\;\text{V/cm}$ ($B_{cr}=m^2c^3/\hbar|e|=4.4\times 10^{13}\;\text{G}$), with $e<0$ and $m$ being the electric charge and the mass of the electron, respectively \cite{Ritus_1985}. A constant and uniform electric field of the order of $E_{cr}$ would be able to provide an electron and a positron with an energy of the order of $mc^2$ over a distance of the order of the Compton wavelength $\lambdabar_c=\hbar/mc$, implying the instability of the vacuum in the presence of such a strong field under electron-positron pair creation. Recent investigations have even shown the possibility of ``breaking'' the vacuum at field amplitudes much smaller than $E_{cr}$ \cite{Cascade}.

Modern laser systems are sources of intense electromagnetic fields and provide a unique ``clean'' environment for testing QED under extreme conditions. The record intensity of $2\times 10^{22}\;\text{W/cm$^2$}$ has been already experimentally achieved \cite{Yanovsky_2008} and intensities of the order of $10^{24}\text{-}10^{25}\;\text{W/cm$^2$}$ are envisaged at the Extreme Light Infrastructure (ELI) \cite{ELI} and at the High Power laser Energy Research facility (HiPER) \cite{HiPER} (the critical field amplitude corresponds to a laser intensity of $I_{cr}=cE_{cr}^2/8\pi=2.3\times 10^{29}\;\text{W/cm$^2$}$). One of the most appealing theoretical predictions of strong-field QED in a (almost-)monochromatic laser field approximated as a plane wave is the occurrence of the intensity-dependent electron ``quasi-momentum'' in the energy-momentum conservation relation \cite{NLCS}. The quasi-(four-)momentum is defined classically as the average (four-)momentum in the plane wave over one laser period. If an electron with initial four-momentum $p^{\mu}=(\epsilon/c,\bm{p})=(\sqrt{\bm{p}^2+m^2c^2},\bm{p})$ collides with a monochromatic, linearly-polarized laser field with electric field amplitude $E$ and angular frequency $\omega$, the quasi-momentum $q^{\mu}$ is given by $q^{\mu}=p^{\mu}+m^2c^2\xi^2n^{\mu}/4(np)$, where $\xi=|e|E/m\omega c$ is the classical intensity parameter, $\omega n^{\mu}$ is the wave four-vector of the laser field and $(np)=n^{\mu}p_{\mu}$ (the space-time metric $\eta^{\mu\nu}=\text{diag}(+1,-1,-1,-1)$ is used) \cite{Landau_b_4_1982}. This leads to the introduction of an effective ``dressed'' mass $m_*=m\sqrt{1+\xi^2/2}$ of the electron in the plane wave such that $q^2=m_*^2c^2$. The mass correction $m_*-m$ depends only on the classical parameter $\xi$ and it is related to the quiver motion of the electron in the oscillating external field \cite{Landau_b_4_1982}. The results of the only experiment so far performed on strong-field QED in a background laser field are reported in \cite{SLAC_Exp} and are consistent with an electron having a ``dressed'' mass $m_*$ in the laser field. More recently the question of the electron dressed mass in pulsed laser fields has been investigated in \cite{Heinzl_2010,Mackenroth_2011}.

In this Letter we show that, due to the interaction of the electron with its own radiation field, the states of the electron in a plane wave are modified in a non-trivial and experimentally testable way. We attain this by starting from the one-loop expression of the mass operator found in \cite{Baier_1976a} and by solving the Schwinger-Dirac equation in the limit $\xi\gg 1$ (available optical lasers already achieve values of $\xi\sim 10^2$ \cite{Yanovsky_2008}). The radiatively-corrected states show that: a) the electron quasi-momentum also contains a pure quantum contribution which does not depend on the classical parameter $\xi$ but only on the quantum nonlinearity parameter $\chi=((np)/m c)(E/E_{cr})$; b) the electron self interaction generates a distinctive spin dynamics, which can be in principle revealed with presently available technology (for semiclassical studies of spin evolution in a plane wave see e.g. \cite{Spin}). Below, units with $\hbar=c=1$ are employed.

In order to describe the evolution of an electron in a plane-wave field, we work in the Furry representation in which the presence of the plane-wave field is accounted for exactly from the beginning \cite{Landau_b_4_1982}. For the sake of definiteness we assume that the plane wave propagates along the positive $z$-direction, that it is linearly polarized along the $x$-direction and that it has an electric field amplitude $E$ and a central angular frequency $\omega$. Thus, it can be described by the four-vector potential $A^{\mu}(\phi)=\psi(\phi)\mathscr{A}^{\mu}$, where $\psi(\phi)$ is an arbitrary function of the phase $\phi$ with $|\psi(\phi)|\lesssim 1$ and with the initial condition $A^{\mu}(\phi\to\pm\infty)=0$, and where $\mathscr{A}^{\mu}=(0,a,0,0)$ with $a=E/\omega>0$ is the polarization four-vector. The laser phase $\phi$ is given by $\phi=(nx)=t-z$, with $n^{\mu}=(1,0,0,1)$ being the unit wave four-vector such that $n^2=(n\mathscr{A})=0$. Since the plane-wave field depends only on the phase $\phi$, it is natural to introduce the ``asymmetric'' light-cone components of a four-vector $v^{\mu}$ as $v^-=v^0-v^3$, $v^+=(v^0+v^3)/2$ and $\bm{v}^{\perp}=(v^1,v^2)$ (note that by definition $x^-=\phi$). In order to take into account also the effects on the electron evolution of the electromagnetic field generated by the electron itself, we have to solve the Schwinger-Dirac equation \cite{Landau_b_4_1982,Baier_1976a}
\begin{equation}
\label{SD_Eq}
(\slashed{\Pi}-m-M)|\Psi\rangle=0,
\end{equation}
where the ``slash'' indicates the contraction with the gamma matrices $\gamma^{\mu}$, where $\Pi^{\mu}=P^{\mu}-eA^{\mu}(\phi)$, with $P^{\mu}$ being the canonical four-momentum operator and where $M$ is the mass operator in the plane-wave field, which includes the electron self interaction  (see Fig. 1).
\begin{figure}
\begin{center}
\includegraphics[width=3cm]{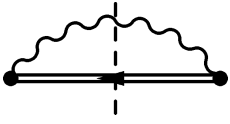}
\end{center}
\caption{Leading-order Feynman diagram of the mass operator in an external plane-wave field. The double solid line indicates the electron propagator in the plane wave (Volkov propagator) and the wiggly line a non-laser photon. The vertical dashed line links the mass operator to the non-linear Compton scattering diagram.}\label{fig:massoperator}
\end{figure}

If self-interaction effects are neglected, then $M=0$ in Eq. (\ref{SD_Eq}) and a complete set of solutions of the resulting equation in coordinate space $\{\gamma^{\mu}[i\partial_{\mu}-eA_{\mu}(\phi)]-m\}\Psi(x)=0$ for an arbitrary plane-wave four-vector potential $A^{\mu}(\phi)$ are the well known Volkov states $\Psi^{(+)}_{V;\sigma}(p,x)$ and $\Psi^{(-)}_{V;\sigma}(p,x)$ with positive and negative energy, respectively \cite{Landau_b_4_1982}. The state $\Psi^{(\pm)}_{V;\sigma}(p,x)$ corresponds to an asymptotically free electron at $\phi\to-\infty$ with on-shell four-momentum $\pm p^{\mu}$ ($p^2=m^2$) and with spin projection $\sigma/2$ ($\sigma=+1$ or $\sigma=-1$) along a given direction in the electron's rest frame (a throughout analysis of the physical meaning of the Volkov states can be found in \cite{Loetstedt_2009}). The Volkov states can be written as $\Psi^{(\pm)}_{V;\sigma}(p,x)=\mathcal{E}(\pm p,x)U^{(\pm)}_{\sigma}(p)/\sqrt{2\epsilon}$, where
\begin{equation}
\label{E}
\mathcal{E}(p,x)=\left(1+\frac{e\slashed{n}\slashed{A}}{2p^-}\right)\text{e}^{-i\left\{(px)+\int_{-\infty}^{\phi}d\phi'\left[\frac{e(pA)}{p^-}-\frac{e^2A^2}{2p^-}\right]\right\}}
\end{equation}
are the so-called Ritus matrices and where the constant spinors $U^{(\pm)}_{\sigma}(p)$ are solutions of the equation $(\slashed{p}\mp m)U_{\sigma}^{(\pm)}(p)=0$ and they are normalized as in \cite{Landau_b_4_1982,Footnote_1}. The expression of the Volkov states indicates that if the electron does not radiate, the spin projection before and after the interaction with the laser field remains unchanged. Furthermore, a Fourier-series expansion of the integrand in the exponent in Eq. (\ref{E}) for a (almost-)monochromatic plane wave shows that the coordinate four-vector $x^{\mu}$ is contracted with the four-vector $q^{\mu}=p^{\mu}-e^2\langle A^2\rangle n^{\mu}/2p^-$, with  $\langle\cdots\rangle$ indicating the average over one laser period, i.e. that the electron propagates in the wave with an ``effective'' four-momentum  $q^{\mu}$, called the quasi-momentum. Interestingly, the difference $q^{\mu}-p^{\mu}$ is of pure classical origin and it does not depend on $\hbar$. For on-shell four-momenta $k^{\mu}$ and $k^{\prime\mu}$ and for four-coordinates $x^{\mu}$ and $x^{\prime\mu}$ such that $x^-=x^{\prime-}$ the Ritus matrices fulfill the three-dimensional completeness and orthogonality relations
\begin{align}
\label{C}
&\int\frac{d^3\accentset{\leadsto}{\bm{k}}}{(2\pi)^3}\mathcal{E}(k,x)\bar{\mathcal{E}}(k,x')=\delta^3(\accentset{\leadsto}{\bm{x}}-\accentset{\leadsto}{\bm{x}}'),\\
\label{P}
&\int \frac{d^3\accentset{\leadsto}{\bm{x}}}{(2\pi)^3}\bar{\mathcal{E}}(k,x)\mathcal{E}(k',x)=\delta^3(\accentset{\leadsto}{\bm{k}}-\accentset{\leadsto}{\bm{k}}'),
\end{align}
where in general $\bar{\mathcal{E}}(k,x)=\gamma^0\mathcal{E}^{\dag}(k,x)\gamma^0$ and where we have introduced the three-dimensional quantities $\accentset{\leadsto}{\bm{x}}=(x^+,\bm{x}^{\perp})$ and $\accentset{\leadsto}{\bm{k}}=(k^-,\bm{k}^{\perp})$. Note that in Eqs. (\ref{C}) and (\ref{P}) all the integrals are from $-\infty$ to $+\infty$ and they also include the negative-energy region with $k^-<0$. Below, however, we will only calculate the radiatively-corrected positive-energy states $\Psi^{(+)}(x)$.

In order to solve Eq. (\ref{SD_Eq}), we observe that in the worst case radiative corrections in a plane wave scale as $\alpha\chi^{2/3}$ at $\chi\gg 1$, with $\alpha=e^2\approx 1/137$ \cite{Ritus_1985}. Therefore, if $\alpha\chi^{2/3}\approx 1$ it is even not possible to write the Schwinger-Dirac equation explicitly, as the mass operator has to be determined in principle at all loops. This prevents at least in practice the determination of the electron states for arbitrary laser and electron parameters. However, available laser systems and electron accelerators allow only for values of $\chi$ such that $\alpha\chi^{2/3}\ll 1$. For example, even in the head-on collision of the most intense available laser field \cite{Yanovsky_2008} with an electron beam of $46.6\;\text{GeV}$ \cite{SLAC_Exp} one obtains $\alpha\chi^{2/3}\approx 0.1$ and it is sufficient to employ the one-loop expression of the mass-operator as calculated in \cite{Baier_1976a}. For the same reason, one can consistently approximate $\slashed{\Pi}\approx m$ in $M$ in Eq. (\ref{SD_Eq}), which considerably simplifies the structure of the mass operator. Moreover, already available optical laser systems allow for values of the parameter $\xi$ much larger than unity. It has been shown that in this ``quasi-static'' limit $\xi\gg 1$ the one-loop mass operator reduces to the analogous quantity in a constant-crossed field $F^{\mu\nu}$, with the substitution $F^{\mu\nu}\to f^{\mu\nu}\psi'(\phi)$, where $f^{\mu\nu}=n^{\mu}\mathscr{A}^{\nu}-n^{\nu}\mathscr{A}^{\mu}$ \cite{Ritus_1985}. We recall that a constant-crossed field $F^{\mu\nu}$ is a constant electromagnetic field such that $F^{\mu\nu}F_{\mu\nu}=F^{\mu\nu}F^*_{\mu\nu}=0$, where $F^{*,\mu\nu}=\epsilon^{\mu\nu\lambda\rho}F_{\lambda\rho}/2$ with $\epsilon^{\mu\nu\lambda\rho}$ being the antisymmetric tensor ($\epsilon^{0123}=1$).

The above Eqs. (\ref{C}) and (\ref{P}) allow to employ the Ritus matrices as a generalized three-dimensional ``Fourier'' transform with e.g. $\Psi(p,\phi)=\int d^3\accentset{\leadsto}{\bm{x}}\bar{\mathcal{E}}(p,x)\Psi(x)$ and so on. This is extremely useful as in the regime of interest here ($\xi\gg 1$ and $\alpha\chi^{2/3}\ll 1$) the one-loop mass operator $M(p,\phi,p',\phi')$ in momentum space for on-shell four-momenta $p^{\mu}$ and $p^{\prime\mu}$ results diagonal \cite{Ritus_1985,Baier_1976a}: $M(p,\phi,p',\phi')=(2\pi)^3\delta(\phi-\phi')\delta^3(\accentset{\leadsto}{\bm{p}}-\accentset{\leadsto}{\bm{p}}')M(p,\phi)$. Therefore, the ``Fourier''-transformed Schwinger-Dirac equation reads
\begin{equation}
\label{SD_Eq_p}
\bigg[i\slashed{n}\frac{\partial}{\partial \phi}+\slashed{p}-m-M(p,\phi)\bigg]\Psi^{(+)}(p,\phi)=0,
\end{equation}
where we have used the fact that for any function $g(\phi)$ it is $\gamma^{\mu}[i\partial_{\mu}-eA_{\mu}(\phi)]\mathcal{E}(p,x)g(\phi)=\mathcal{E}(p,x)[\slashed{p}g(\phi)+i\slashed{n}\,g'(\phi)]$. From the expression of the mass operator in a constant crossed field ($\psi'(\phi)=1$) in Eq. (32) on pg. 591 in \cite{Ritus_1985} with $i\slashed{p}=-m$ there, one can write $M(p,\phi)$ as $M(p,\phi)=m\{C_0(p,\phi)+D_0(p,\phi)\Sigma+(m/p^-)[C_1(p,\phi)+D_1(p,\phi)\Sigma]\slashed{n}\}$, where
\begin{align}
\label{r_1}
C_j(p,\phi)&=-\alpha\int_0^{\infty}\frac{dl}{2\pi}\frac{(j-3)l^2+(-8)^jl+(-6)^j}{3(1+l)^3\rho}\frac{df(\rho)}{d\rho},\\ 
\label{r_2}
D_j(p,\phi)&=-\alpha\int_0^{\infty}\frac{dl}{2\pi}\frac{(-2)^jl+(-1)^j2}{(1+l)^3}\frac{f(\rho)}{\sqrt{\rho}},
\end{align}
with $j\in\{0,1\}$, $\rho=(l/\chi(\phi))^{2/3}$, $\chi(\phi)=ap^-\psi'(\phi)/mE_{cr}$ and $f(z)=\pi[\text{Gi}(z)+i\text{Ai}(z)]$ \cite{NIST_b_2010} and where
\begin{equation}
\Sigma=\frac{1}{ap^-}\gamma^5f^{*,\mu\nu}p_{\mu}\gamma_{\nu},
\end{equation}
with $\gamma^5=-i\gamma^0\gamma^1\gamma^2\gamma^3$ is the spin operator. The eigenvalues of the operator $\Sigma$ are (twice) the projection of the electron spin along the direction of the magnetic field of the plane wave in the initial rest frame of the electron. Note that the mass operator $M(p,\phi)$ is renormalized as the corresponding quantity in vacuum $M_0(p)$ because $M(p,\phi)=[M(p,\phi)-M_0(p)]+M_0(p)$ and the quantity $M(p,\phi)-M_0(p)$ is found to be regular \cite{Baier_1976a}. Since the two commutators $[\slashed{p},\Sigma]$ and $[\slashed{n},\Sigma]$ vanish, we can choose the solutions $\Psi^{(+)}_{\sigma}(p,\phi)$ of Eq. (\ref{SD_Eq_p}) as eigenstates of $\Sigma$: $\Sigma\Psi^{(+)}_{\sigma}(p,\phi)=\sigma\Psi^{(+)}_{\sigma}(p,\phi)$, with $\sigma=\pm 1$. In this way $M(p,\phi)\Psi^{(+)}_{\sigma}(p,\phi)=M_{\sigma}(p,\phi)\Psi^{(+)}_{\sigma}(p,\phi)$, where
\begin{equation}
M_{\sigma}(p,\phi)=m\left[R_{0,\sigma}(p,\phi)+\frac{m}{p^-}R_{1,\sigma}(p,\phi)\slashed{n}\right],
\end{equation}
with $R_{j,\sigma}(p,\phi)=C_j(p,\phi)+D_j(p,\phi)\sigma$. The matrix structure of Eq. (\ref{SD_Eq_p}) with the given expression of $M_{\sigma}(p,\phi)$, suggests to write the (positive-energy) solution $\Psi^{(+)}_{\sigma}(p,\phi)$ as $\Psi^{(+)}_{\sigma}(p,\phi)=\exp[(m/2p^-)R_{0,\sigma}(p,\phi)\slashed{n}]\Phi^{(+)}_{\sigma}(p,\phi)$, with the spinor $\Phi^{(+)}_{\sigma}(p,\phi)$ being a solution of the equation
\begin{equation}
\label{SD_Eq_p_1}
\bigg[i\slashed{n}\frac{\partial}{\partial\phi}+\slashed{p}-m-\frac{m^2}{p^-}\slashed{n}\sum_{j=0}^1R_{j,\sigma}(p,\phi)\bigg]\Phi_{\sigma}^{(+)}(p,\phi)=0,
\end{equation}
where terms quadratic in $\alpha$ have been consistently neglected. The solution of this equation with on-shell four-momentum $p^{\mu}$ is given by $\Phi_{\sigma}^{(+)}(p,\phi)=\exp\{-i(m^2/p^-)\int_{-\infty}^{\phi}d\phi'[R_{0,\sigma}(p,\phi')+R_{1,\sigma}(p,\phi')]\}U^{(+)}_{\sigma}(p)/\sqrt{2\epsilon}$, with $\Sigma U^{(+)}_{\sigma}(p)=\sigma U^{(+)}_{\sigma}(p)$. Therefore, the final result for the radiatively-corrected state $\Psi^{(+)}_{R,\sigma}(p,x)=\mathcal{E}(p,x)\Psi^{(+)}_{\sigma}(p,\phi)$ in configuration space can be written in a very compact way as $\Psi^{(+)}_{R,\sigma}(p,x)=\mathcal{E}_{R,\sigma}(p,x)U^{(+)}_{\sigma}(p)/\sqrt{2\epsilon}$, where the radiatively-corrected Ritus matrices
\begin{equation}
\label{E_1}
\begin{split}
\mathcal{E}_{R,\sigma}(p,x)=&\left\{1+\frac{m\slashed{n}}{2p^-}\left[\frac{e\slashed{A}}{m}+R_{0,\sigma}(p,\phi)\right]\right\}\text{e}^{iS_{R,\sigma}(p,x)},
\end{split}
\end{equation}
with the radiatively-corrected electron ``action''
\begin{equation}
\label{S_1}
\begin{split}
S_{R,\sigma}(p,x)=&-(px)-\int_{-\infty}^{\phi}d\phi'\left\{\frac{e(pA)}{p^-}-\frac{e^2A^2}{2p^-}\right.\\
&+\frac{m^2}{p^-}[R_{0,\sigma}(p,\phi')+R_{1,\sigma}(p,\phi')]\bigg\}
\end{split}
\end{equation}
have been introduced. The above expression of the state $\Psi^{(+)}_{R,\sigma}(p,x)$ is the main result of the Letter. The expression of $\mathcal{E}_{R,\sigma}(p,x)$ in Eq. (\ref{E_1}) has to be intended as being valid up to first-order in $\alpha$. However, the integral on $\phi'$ of the function $R_{0,\sigma}(p,\phi')+R_{1,\sigma}(p,\phi')$ in $S_{R,\sigma}(p,x)$ may become large and it has to be taken into account exactly. The structure of Eq. (\ref{S_1}) allows us in the case of a (almost-)monochromatic laser field to introduce for the first time the radiatively-corrected quasi-momentum
\begin{equation}
Q^{\mu}=q^{\mu}+\frac{m^2}{p^-}[\langle C_0\rangle(p)+\langle C_1\rangle(p)]n^{\mu},
\end{equation}
where we used the fact that the average values of $D_j(p,\phi)$ vanish for such a pulse (see Eqs. (\ref{r_1}) and (\ref{r_2})). However, as we will see below, the total integral of the quantities $D_j(p,\phi)$ over a few-cycle pulse does not necessarily vanish, even though for short pulses the introduction of the concept of ``quasi-momentum'' is not straightforward (see also \cite{Heinzl_2010,Mackenroth_2011}). We note that the radiatively-corrected part $\delta Q^{\mu}=Q^{\mu}-q^{\mu}$ of the quasi-momentum depends on the quantum parameter $\chi$ and it has no classical counterpart. In fact, by calculating its asymptotic value in the classical limit $\chi\to 0$, one finds that the leading contribution scales as $\alpha\chi^2$, which is proportional to $\hbar$. On the other hand, Eqs. (\ref{r_1}) and (\ref{r_2}) show that $\delta Q^{\mu}$ contains in general a (negative) imaginary part as the electron states in the laser field are ``unstable'': an electron in a laser field emits photons and in this way ``decays'' (see also Fig. 1). As a check of our results, we have ensured that if radiative corrections are neglected then $\Psi^{(\pm)}_{R,\sigma}(p,x)$ reduces to $\Psi^{(\pm)}_{V,\sigma}(p,x)$, and that the radiative correction $\delta m_*^2\equiv Q^2-q^2=2m^2[\langle C_0\rangle(p)+\langle C_1\rangle(p)]$ to the square of the electron's effective mass is in agreement with the results in \cite{Ritus_1970,Baier_1976b} (see also Eq. (54) on pg. 595 in \cite{Ritus_1985}).

The dependence of $S_{R,\sigma}(p,x)$ in Eq. (\ref{S_1}) on $\sigma$ indicates that the electron self-interaction affects the electron spin dynamics. In the following we suggest a possible experimental setup to measure this effect. An on-shell electron with initial four-momentum $p^{\mu}=(\epsilon,0,0,-p)$ and spin polarization in its rest frame along the positive $x$-direction ($\sigma_x=+1$), head-on collides with a laser beam linearly polarized along the $x$-direction, with electric field amplitude $E$ and central angular frequency $\omega$ (wavelength $\lambda=2\pi/\omega$). The electron state in the laser field is then $\Psi^{(+)}_{R,\sigma_x=+1}(p,x)$ corresponding asymptotically to the free state $\Psi^{(+)}_{F,\sigma_x=+1}(p,x)\equiv\exp(-i(px))U^{(+)}_{\sigma_x=+1}(p)/\sqrt{2\epsilon}$ at $\phi\to -\infty$. Since the magnetic field of the laser lies along the $y$-direction, it is convenient to employ the eigenstates $\Psi^{(+)}_{R,\sigma_y}(p,x)$  ($\sigma_y=\pm 1$) of the spin-projection along this direction to expand the state $\Psi^{(+)}_{R,\sigma_x=+1}(p,x)$. Eqs. (\ref{E_1}) and (\ref{S_1}) show that after the interaction with the laser field the component with a given $\sigma_y$ acquires a phase factor of the form $\exp(-i\Phi_{\sigma_y})$, with $\Phi_{\sigma_y}=\Phi_0+\sigma_y \Phi_s$, and the probabilities $P_{\uparrow/\downarrow}$ that the electron spin points along the positive/negative $z$-direction, i.e. that the asymptotically free electron state at $\phi\to+\infty$ is $\Psi^{(+)}_{F,\sigma_z}(p,x)$ with $\sigma_z=\pm 1$, are given by
\begin{equation}
\label{P_ud}
P_{\uparrow/\downarrow}=\frac{\text{e}^{2\text{Im}(\Phi_0)}}{2}\big[\cosh(2\text{Im}(\Phi_s))\mp\sin(2\text{Re}(\Phi_s))\big],
\end{equation}
respectively. Note that since the electron is now ``unstable'', the sum $P_{\uparrow}+P_{\downarrow}$ becomes less than unity.
\begin{figure}
\begin{center}
\includegraphics[width=7cm]{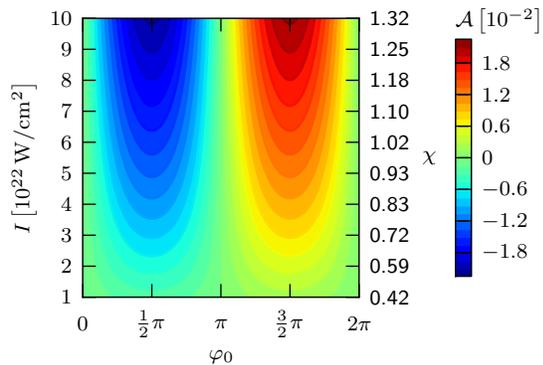}
\end{center}
\caption{(Color online) Asymmetry $\mathcal{A}$ for numerical parameters given in the text below Eq. (\ref{P_ud}).}
\end{figure}
We introduce the asymmetry $\mathcal{A}\equiv(P_{\uparrow}-P_{\downarrow})/(P_{\uparrow}+P_{\downarrow})=-\sin(2\text{Re}(\Phi_s))/\cosh(2\text{Im}(\Phi_s))$ as a convenient observable. In the following numerical calculations we employ a laser profile given by $\psi'(\phi)=\omega\sin^2(\omega\phi/2N)\sin(\omega\phi+\varphi_0)$ if $\phi\in[0,N\lambda]$ and zero otherwise. In Fig. 2 we show the asymmetry $\mathcal{A}$ as a function of the laser intensity $I$ and of the laser carrier-envelope phase (CEP) $\varphi_0$ for the following numerical parameters: $\omega=1.55\;\text{eV}$, $N=3$ (corresponding to a pulse duration $\tau\approx 8\;\text{fs}$) and $\epsilon=500\;\text{MeV}$. For these parameters it is $|\Phi_s|\ll 1$ and $\mathcal{A}\approx-2\text{Re}(\Phi_s)=-2(m^2/p^-)\text{Re}\int_0^{N\lambda}d\phi[D_0(p,\phi)+D_1(p,\phi)]$. Since for long pulses a relatively large amount of electrons will ``decay'' by radiating, it is convenient to employ here rather short pulses (we have ensured that the asymmetry becomes insensitive to the CEP for longer pulses). Phase-stabilized laser pulses with $\tau\sim 5\;\text{fs}$ and $I\gtrsim 10^{22}\;\text{W/cm$^2$}$ are experimentally envisaged \cite{PFS}. At an intensity of $I_0=4\times 10^{22}\;\text{W/cm$^2$}$, for example, we find $\text{exp}[2\text{Im}(\Phi_0)]\approx 9\times 10^{-4}$ and a maximal asymmetry $\mathcal{A}_0\sim 1\;\%$ (see Fig. 2), such as those already measured even at electron energies $\sim 1\;\text{GeV}$ \cite{Escoffier_2005}. Polarized ultra-relativistic electron beams with $\mathcal{N}\approx 10^{10}$ electrons, a spot area $\approx 1.7\;\text{$\mu$m}\times 0.75\;\text{$\mu$m}$ and a length $l_e$ of about $0.5\;\text{mm}$ have been produced \cite{FFTB}. Assuming a Gaussian laser beam focused to one wavelength (spot radius $w_0=\lambda$ and Rayleigh length $l_r=\pi w_0^2/\lambda=\pi\lambda$) \cite{Saleh_b_1991}, about $\mathcal{N}^*\sim \mathcal{N}\times\text{exp}[2\text{Im}(\Phi_0)]\times 2l_r/l_e\sim 10^5$ electrons pass through the strong-field region without radiating. Thus, the absolute difference of the expected electrons with opposite spin is $\sim \mathcal{A}_0\times \mathcal{N}^*\sim 10^3$. Note also that in the above example the transverse excursion of an electron in the field is approximately $\lambda m\xi/\epsilon\approx 0.14\,\lambda$, i.e. much smaller than $w_0$ \cite{Landau_b_2_1975}. Finally, the relatively weak dependence of the asymmetry on the laser intensity at $I\gtrsim 4\times 10^{22}\;\text{W/cm$^2$}$ renders our results sufficiently insensitive to possible fluctuations of the laser intensity itself.

We gratefully acknowledge useful discussions with K. Z. Hatsagortsyan and C. H. Keitel.
%
%

\end{document}